\documentclass[12pt,preprint]{aastex}
\slugcomment{\it Accepted by ApJ Letters}
\lefthead{Griffith et al.}
\righthead{WISE: Blue Compact Dwarf Galaxies}

\usepackage{graphicx} 
\usepackage{amssymb}
\usepackage{amsmath}
\usepackage{anysize}
\usepackage{rotating}
\usepackage{multirow}
\usepackage{float}

\def\eg{{e.g.,~}}
\def\etal{{et al.}}

\def\spose#1{\hbox to 0pt{#1\hss}}
\def\simlt{\mathrel{\spose{\lower 3pt\hbox{$\mathchar"218$}}
     \raise 2.0pt\hbox{$\mathchar"13C$}}}
\def\simgt{\mathrel{\spose{\lower 3pt\hbox{$\mathchar"218$}}
     \raise 2.0pt\hbox{$\mathchar"13E$}}}

\def\wise{{\it WISE}}
\def\wzero{{W0801+26}}
\def\wone{{W1702+18}}
\def\sbs{{SBS~0335-052E}}
\def\zwicky{{I~Zw~18}}

\begin{document}

\title{WISE Discovery of
Low Metallicity \\ Blue Compact Dwarf Galaxies}

\author{R\textsc{oger} L. G\textsc{riffith}\altaffilmark{1},
C\textsc{hao}-W\textsc{ei} T\textsc{sai}\altaffilmark{1}, 
D\textsc{aniel} S\textsc{tern}\altaffilmark{2}, 
A\textsc{ndrew} B\textsc{lain}\altaffilmark{3},
P\textsc{eter} R. M. E\textsc{isenhardt}\altaffilmark{2},
F\textsc{iona} H\textsc{arrison}\altaffilmark{4}, 
T\textsc{homas} H. J\textsc{arrett}\altaffilmark{1},
K\textsc{ristin} M\textsc{adsen}\altaffilmark{4},
S\textsc{pencer} A. S\textsc{tanford}\altaffilmark{5},
E\textsc{dward} L. W\textsc{right}\altaffilmark{6},
J\textsc{ingwen} W\textsc{u}\altaffilmark{2},
Y\textsc{anling} W\textsc{u}\altaffilmark{1} \\
\& L\textsc{in} Y\textsc{an}\altaffilmark{1}}

\altaffiltext{1}{Infrared Processing and Analysis Center, California Institute of Technology, Pasadena, CA 91125}
\altaffiltext{2}{Jet Propulsion Laboratory, California Institute of Technology, 4800 Oak Grove Dr., Pasadena, CA 91109}
\altaffiltext{3}{Department of Physics and Astronomy, University of Leicester, LE1 7RH Leicester, UK}
\altaffiltext{4}{Space Radiation Laboratory, California Institute of Technology, Pasadena, CA 91125}
\altaffiltext{5}{Department of Physics, University of California Davis, One Shields Ave., Davis, CA 95616}
\altaffiltext{6}{Astronomy Department, University of California Los Angeles, P.O. Box 951547, Los Angeles, CA 90095}


\begin{abstract}

We report two new low metallicity blue compact dwarf galaxies (BCDs),
WISEP J080103.93+264053.9 (hereafter \wzero) and WISEP~J170233.53+180306.4
(hereafter \wone), discovered using the {\it Wide-field Infrared
Survey Explorer (WISE)}.  We identified these two BCDs from their
extremely red colors at mid-infrared wavelengths, and obtained
follow-up optical spectroscopy using the Low Resolution Imaging
Spectrometer on Keck~I.  The mid-infrared properties of these two
sources are similar to the well studied, extremely low metallicity
galaxy \sbs\ .  We determine metallicities of $12 + \log{\rm
(O/H)} = 7.75$ and 7.63 for \wzero\ and \wone, respectively, placing
them amongst a very small group of very metal deficient galaxies ($Z
\le 1/10\, Z_\sun$).  Their $> 300$ \AA\ H$\beta$ equivalent widths,
similar to \sbs, imply the existence of young ($< 5$ Myr)
star forming regions.  We measure star formation rates of 2.6 and
10.9 $M_\sun$ yr$^{-1}$ for \wzero\ and \wone, respectively.  These
BCDs, showing recent star formation activity in extremely low
metallicity environments, provide new laboratories for studying
star formation in extreme conditions and are low-redshift analogs
of the first generation of galaxies to form in the universe.  Using
the all-sky \wise\ survey, we discuss a new method to identify
similar star forming, low metallicity BCDs.

\end{abstract}

\section{Introduction}

In the local universe, we observe a class of galaxies known as blue
compact dwarf galaxies (BCDs) which show signatures of strong,
recent and ongoing star-formation activity in low metallicity environments.
These galaxies, which appear to be experiencing their first major burst
of star formation, are characterized as generally having compact
sizes ($\le 1$ kpc), low luminosities ($M_B \ge -18$), strong
emission line spectra, blue optical colors ($U - B \sim -0.6$), and
their metal distribution peaks at one-tenth solar \citep[$Z = 1/10\
Z_\odot$;][]{kunth2000}.  Typical H$\beta$ equivalent widths range
from 30 to 200 \AA, implying a burst of star formation no older
than a few Myr \citep{kunth99}.  BCDs offer pristine laboratories
from which to study star formation processes, enrichment mechanisms
for the interstellar medium, and the star formation history of
galaxies in low metallicity and extremely dusty environments.  In
particular, BCDs are likely good analogs of the first generation
of galaxies to form in the universe, but are at redshifts accessible
to detailed study \cite[e.g.,][]{meier81}.  For a comprehensive review of BCDs and the most
metal-poor galaxies, see \cite{kunth2000}.

Although discovered as early as the 1960's, BCDs and extremely low
metallicity galaxies have remained an elusive population. The
difficulty in finding low metallicity BCDs can be viewed as stemming
from several factors.  First, they have low luminosities, generally
much fainter than $L^*$.  Second, at the lowest metallicities, below
0.1 solar, the major coolant species shift from [\ion{O}{3}] to H
and He \citep{kunth86}, resulting in lower [{\ion{O}{3}]
$\lambda$5007/H$\beta$ ratios.  This made such sources less readily
identified in the previous generation of objective prism surveys
\citep{kunth86}, though recent fiber spectroscopy surveys have less significant
biases. Finally, they are rare. The most abundant samples of
BCDs have been acquired by the Sloan Digital Sky Survey (SDSS) which
has spectroscopically identified hundreds of BCDs at redshifts of
a few tenths \citep[\eg][]{izotov06}, implying a surface density
of 0.1 BCD per deg$^2$ to the limits of the SDSS.  However, SDSS
rarely finds the most extreme metal poor systems, galaxies with $Z
\simlt 1/10\ Z_\odot$.  For example, \cite{knia03} estimate 0.004
deg$^{-2}$ for extremely metal poor galaxies [XMPGs; $12+\log{\rm (O/H)}
\le 7.65$] for $r \le 17.77$~mag. 

The two lowest metallicity BCDs known to date are \sbs\ \citep{izotov90,
mel92, izotov97, thuan97, thuan99, dale01, houck04} and \zwicky\
\citep{zwicky66, lauqu73, thuan81, david85, dufour90, kunth94,
hirash04,Wu07}, both of which have been extensively studied across
the electromagnetic spectrum.  Though the metallicities for these
two systems are very similar, $12 + \log{\rm (O/H)} < 7.3$ or $Z <
1/40\, Z_\sun$, these galaxies exhibit very different morphological
and physical properties.  \cite{hirash04} investigated these
differences and suggest that they are best understood if two different
modes of star formation are considered, an ``active mode'' and a
``passive mode''. \sbs\ is an active-mode BCD while \zwicky\ is a
passive-mode BCD.  In the \citet{hirash04} framework, active BCDs
host super star clusters (SSCs) and are characterized by compact
star formation regions (radius $\le 50$ pc), high gas densities
($\ge 500$ cm$^{-3}$), rich H$_2$ content, large dust optical depth,
and high dust temperature.  Passive BCDs, in contrast, have more
diffuse star forming regions ($\ge 100$ pc), lower gas density ($\le
100$ cm$^{-3}$), cooler dust, and lack SSCs and large amounts of
H$_2$.

\citet{hirash04} describe how the compactness of the star forming
regions can simultaneously explain most of the observed differences
between BCDs.  Compact star-forming regions will rapidly become 
dusty from Type II supernovae, and this dust will be effective at 
reprocessing photons into the infrared.  In particular, the gas free-fall 
time scale in such systems is less
than 5~Myr, which leads to run-away star formation and the efficient
creation of large quantities of dust.  In contrast, the dynamical
time of diffuse star forming regions, characteristic of passive
BCDs, is longer than $10^7$~yr.  Such regions thus have lower star
formation rates and are less infrared luminous.
\citet{hirash04} suggest that the physical state of the gas might
be driven by the size and spatial distribution of dust grains.

\cite{wu08} show that the infrared luminosities of these prototypical
BCDs differ by two orders of magnitude, with \sbs\ being $\sim
100$ times more infrared luminous than \zwicky. The mid-infrared
emission of similar low metallicity BCDs originates mainly from hot
(200 -- 1500~K) dust, causing active BCDs to have significantly red
colors at mid-infrared wavelengths.  For example, \cite{houck04}
finds that the spectrum of \sbs\ peaks at $\sim$ 28 $\mu$m.
\cite{dale2001} model the infrared emission of \sbs \ and conclude
that it can be characterized as having two dust components, a warm
($\sim$ 80 K) and a hot ($\sim$ 210 K) component. The prevalence 
of hot dust  in active BCDs suggests that searching for the most extreme 
low metallicity, star-forming BCDs at mid-infrared wavelengths may be fruitful.

This {\it Letter} reports the discovery of two new low metallicity
BCDs discovered by the {\it Wide-field Infrared Survey Explorer}
\citep[\wise;][]{wright10}. Launched on 2009 December 14, \wise\ 
completed its first coverage of the entire sky on 2010 July 17, 
obtaining at least  8 exposures per sky position in four passbands, 
3.4, 4.6, 12 and 22 $\mu$m (W1,W2,W3 and W4). The corresponding 
$5\sigma$ point source sensitivities in unconfused regions are 
better than 0.08, 0.11, 1 and 6 mJy, respectively, and the point spread
function FWHM values for the four WISE bands are 6\farcs1, 6\farcs4, 6\farcs5 and 12\farcs0, 
respectively \citep{wright10}. Focussing on two early examples of
\wise-identified, low metallicity BCDs, we discuss the physical
properties of these two new BCDs and place them within the context
of recent BCD literature.  Tsai et al. (in prep.) presents a
larger sample of several dozen newly identified BCDs and discusses
the selection technique in more detail.  We assume a solar
metallicity ($Z_\sun$) of $12 + \log{\rm (O/H)} = 8.96$ \citep{AP01}
in this paper.  Magnitudes are reported in
the Vega system and, where necessary, we adopt the concordance
cosmology with $\Omega_M = 0.3$, $\Omega_\Lambda = 0.7$
and $H_0 = 70\, {\rm\,km\,s^{-1}\,Mpc^{-1}}$.

\section{BCD Selection and Keck Spectroscopy}

We identified WISEP~J080103.93+264053.9 (hereafter \wzero) and
WISEP~J170233.53+180306.4 (hereafter \wone) from their extreme
mid-infrared colors and relatively bright optical flux densities
($B < 19.5$, from the USNO-B1 catalog).  The mid-infrared photometry
was acquired from the \wise\ Preliminary Release Source Catalog
(thus the ``P'' designation in the \wise\ nomenclature).  The simultaneous
multi-wavelength \wise\ observations of \wzero\ and \wone\ were
carried out on UT 2010 April 12 and UT 2010 March 2, respectively,
and provided total exposure times of $\sim$ 100 and 120 sec,
respectively.  For details on the \wise\ processing in the Preliminary
Release, see the WISE Preliminary Release Explanatory Supplement\footnote{
{\tt http://wise2.ipac.caltech.edu/docs/release/prelim/index.html.}}. \wise\
images of the two new BCDs are presented in Fig.~{\ref{fig:images}}.
For comparison, we also show \wise\ images of the two prototypical
low metallcity BCDs, \sbs\ and \zwicky.  Photometry for all
four sources is presented in Table~1.  The new BCDs have unique and
easily identified mid-infrared colors, similar to \sbs\ .   In
contrast, \zwicky\ is unremarkable in the \wise\ data.  None of the
sources are resolved by the 6\arcsec\, \wise\ beam.  The signal-to-noise
ratios of \sbs\ and the new BCDs are $>$ 25 in all four \wise\
bands.

Noting their unusual \wise\ colors and relatively bright optical
magnitudes, we obtained optical spectroscopic follow-up
observations of \wzero\ and \wone\ during twilight using the dual-beam
Low Resolution Imaging Spectrometer \citep[LRIS;][]{oke95} on the
Keck~I telescope.  We observed \wone\ on UT 2010 March 12 and \wzero\ on UT 2010
November 8.  The conditions were photometric on both nights,
and both observations used the 1\farcs5 wide longslit, the 5600
\AA\ dichroic and the 400 $\ell\, {\rm mm}^{-1}$ grating on the red
arm of the spectrograph (blazed at 8500 \AA; resolving power
$R \equiv \lambda / \Delta \lambda \sim  700$ for objects filling
the slit).  The March 2010 observations used the 600 $\ell\, {\rm
mm}^{-1}$ grism on the blue arm (blazed at 4000 \AA; 
$R \sim 1000$), while the November 2010 observations
used the slightly lower resolution 400 $\ell\, {\rm mm}^{-1}$ grism
(blazed at 3400 \AA; $R \sim 600$).  A single
300 second exposure was obtained of each source during the observing
runs. We processed the data using standard procedures, flux
calibrated the spectra using observations of standard stars from
\citet{massey90}, and corrected for Galactic extinction using the dust maps of \citet{schlegel98}.  

The final, reduced spectra show a multitude of extremely high
equivalent width, narrow emission lines at low redshift and are
presented in Fig.~{\ref{fig:spectra}}.  We have used the ratio of
H$\alpha$ to H$\beta$ and H$\gamma$ to H$\delta$ emission line strengths to derive and correct
for intrinsic dust extinction assuming Case~B recombination and the
extinction law of \citet{CCM89} with $R_V = 3.1$.  For \wzero, the Balmer line ratios are 
close to the extinction-free ratios of Case~B, implying no significant extinction is observed.  For
\wone, we find an intrinsic extinction of $A_V = 0.18$.  The derived
redshifts are presented in Table~1, while Table~2 presents the
extinction-corrected line flux measurements and errors obtained
using the {\tt SPECFIT} package within IRAF. Equivalent widths from
H$\beta$ are also given in Table~2.

\section{Results}

\subsection{Mid-Infrared Properties}

Fig.~{\ref{fig:images}} shows \wise\ color images of \wzero\
and \wone, as well as of prototypical BCDs \sbs\ and \zwicky.  In
the framework of \citet{hirash04}, \sbs\ is an active BCD, while
\zwicky\ is passive BCD.  Most Galactic stars and low redshift
galaxies are dominated by a Rayleigh-Jeans spectrum at the wavelengths
observed by \wise, providing blue colors in the color mapping shown
in Fig.~{\ref{fig:images}}. In contrast, active BCDs contain substantial 
quantities of dust in a dense cloud, which give them unusual and 
readily-identified red colors at near- and mid-infrared wavelengths.  
BCDs like \zwicky, on the other hand, do not contain hot dust, and thus are undistinguished in the 
\wise\ bands (see Table~1).

As discussed in Yan \etal\ (in prep.), sources with red colors
across the first two \wise\ bands, $W1 - W2 \ge 2$, are relatively
rare.  For example, the COSMOS field has only one source with such
red \wise\ colors in a field slightly larger than 1 deg$^2$ (Stern
\etal, in prep.).  The coolest brown dwarfs, with spectral types
later than T5, have red $W1 - W2$ colors due to methane absorption
at 3.3 $\mu$m \citep[\eg][]{burgasser2011, kirkpatrick2011, mainzer2011}
and are quite faint at optical wavelengths.  The \wise\ team has
also been pursuing a population of galaxies with red $W2 - W4$
colors which appear to be dominated by heavily obscured AGN at $z
\sim 2$ (\eg Eisenhardt \etal, in prep.).  Such galaxies are also
quite faint optically. In contrast, as shown in Table~1, low
redshift BCDs have extremely red \wise\ colors but are bright at
optical wavelengths, providing for their easy identification and
follow-up.  Tsai \etal\ (in prep.) discusses the \wise\ selection
of BCDs in more detail.

\subsection{Spectroscopic Properties and Metallicity}

The Keck spectra of \wzero\ and \wone, presented in
Fig.~{\ref{fig:spectra}}, show a multitude of strong hydrogen
and oxygen emission lines as well as weak stellar continuum.  These
spectra are characteristic of sources whose optical emission
is dominated by \ion{H}{2} regions.  Key line measurements for
\wzero\ and \wone\ are presented in Table~2.  Standard diagnostic
diagrams, such as comparing [\ion{O}{3}]/H$\beta$ to
[\ion{N}{2}]/H$\alpha$, imply that the emission lines for both new
\wise-selected BCDs are driven by star formation processes rather
than by AGN activity \citep[\eg][]{baldwin81,kewly01}.

One of the most striking features of the spectra presented in
Fig.~{\ref{fig:spectra}} is the weakness of the [\ion{N}{2}] lines
relative to the H$\alpha$ line that they flank.  This immediately
suggests that both galaxies have extremely low metallicities.
Following the prescriptions of \cite{izotov06}, we derive electron
temperatures $T_e$([\ion{O}{3}]) $\sim 1.7 \times 10^4$~K for \wzero \
and $\sim 1.9 \times 10^4$~K for \wone, from their [\ion{O}{3}] line strengths at 4363, 4959 and
5007 \AA.  Based on the relative strengths of the Balmer lines to
these oxygen lines, we then derive total heavy element abundances
and metallicities of $12 + \log{\rm(O/H)} = 7.75^{+0.04}_{-0.17}$ and $12 + \log{\rm(O/H)} = 7.63 \pm 0.06$ for \wzero\ and \wone, respectively, where these errors were derived using a Monte
Carlo method with 10,000 simulations.

Using the H$\alpha$ and 22 $\mu$m flux densities we measure star
formation rates (SFRs) as prescribed in \cite{calzetti07}. We find
SFR = 2.6 and 10.9 $M_\sun$ yr$^{-1}$ for \wzero\ and \wone,
respectively, which imply that these two BCDs are currently undergoing
very extreme bursts of star formation activity. In addition, the
H$\beta$ equivalent widths $>$ 300 \AA\ for \wzero\ and \wone\ imply
that the ages of the stellar populations are less than 5 Myr
\citep{schaerer98}.

\section{Discussion and Summary}

The low metallicity and mid-infrared dominated spectral energy
distributions of these newly identified, \wise-selected galaxies
place them in a rare class of active, low metallicity BCDs.  We
compare the infrared properties of these BCDs with the infrared
properties of known BCDs and find that the majority of low metallicity
BCDs in the literature are several magnitudes fainter at 12 $\mu$m
($W3$) than the two new sources. The properties of the two new BCDs
are strikingly similar to one of the most well known BCD, \sbs,
suggesting that the mechanisms responsible for 
\sbs\ might be the same mechanism at work in \wzero\ and \wone.

In Figure 3 we show \wise\ $W1 - W2$ color versus metallicity, $12 +
\log{\rm (O/H)}$, for different samples of BCDs.  
After visual inspections to remove galaxies that clearly are
not BCDs, we include 14 BCDs
from \cite{wu08}, 173 from \cite{izotov06},
three from \cite{knia03}, \sbs, \zwicky, HS 0822+3542 \citep{izotov06}, HS 2236+1344 \citep{izotov2007}
and our two newly discovered BCDs. With the exception of \zwicky\
and HS~0822+3542, we observe a slight correlation between $W1-W2$
and metallicity for these samples, with $W1-W2$ color becoming
increasingly redder with decreasing metallicity.  We apply the
Spearman's rank correlation coefficient test on the  \cite{izotov06} sample and
recover a Spearman coefficient of $\rho = -0.36$.  For a sample of
that size, this indicates a noticeable correlation, deviating from
the null-hypothesis at the 4$\sigma$ level.

Since polycyclic aromatic hydrocarbon (PAH) emission is largely
absent in low-metallicity environments \citep{Engelbracht08}, the
mid-infrared spectra of BCDs is dominated by thermal emission 
with a minor contribution from nebula continuum \citep{smith09}.
Thus, the 3.4\,$\mu$m to 4.6\,$\mu$m flux ratio is a relatively
clean temperature indicator for the hot dust heated by young
stars.  At the lowest metallicitities, $12 + \log{\rm (O/H)} < 7.8$,
BCDs seem to group into two different color regions, $W1-W2 \sim
2.0$ and $\sim 0.5$. This suggests two distinct star-formation
processes can be dominant, supporting the passive and active BCD
modes suggested by \citet{hirash04}.

We are currently conducting follow-up observations of many more
promising candidates discovered using the extreme infrared properties
exhibited by these two sources and \sbs.  These results will be
presented in a follow-up paper (Tsai \etal, in prep.).  We expect
that \wise\ will significantly increase the number of low metallicity
BCDs known, perhaps finding the most pristine example in the local
universe.  Such studies will allow us to construct a statistical
sample from which to study and characterize the BCD population. The
\wise\ BCDs will be a unique sample of mid-infrared bright, low
metallicity galaxies for studying dust grain formation in low
metallicity environments, and will help us to understand the thermal
dust emission of high-redshift, starburst galaxies.

\acknowledgements

The authors thank the anonymous referee for timely and beneficial
comments that have improved the manuscript.  This publication makes
use of data products from the {\it Wide-field Infrared Survey
Explorer}, which is a joint project of the University of California,
Los Angeles, and the Jet Propulsion Laboratory, California Institute
of Technology, funded by the National Aeronautics and Space
Administration.  Some of the data presented herein were obtained
at the W.M. Keck Observatory, which is operated as a scientific
partnership among the California Institute of Technology, the
University of California and the National Aeronautics and Space
Administration. The Observatory was made possible by the generous
financial support of the W.M. Keck Foundation. The authors also
wish to recognize and acknowledge the very significant cultural
role and reverence that the summit of Mauna Kea has always had
within the indigenous Hawaiian community; we are most fortunate to
have the opportunity to conduct observations from this mountain.

\begin{figure}
\plotone{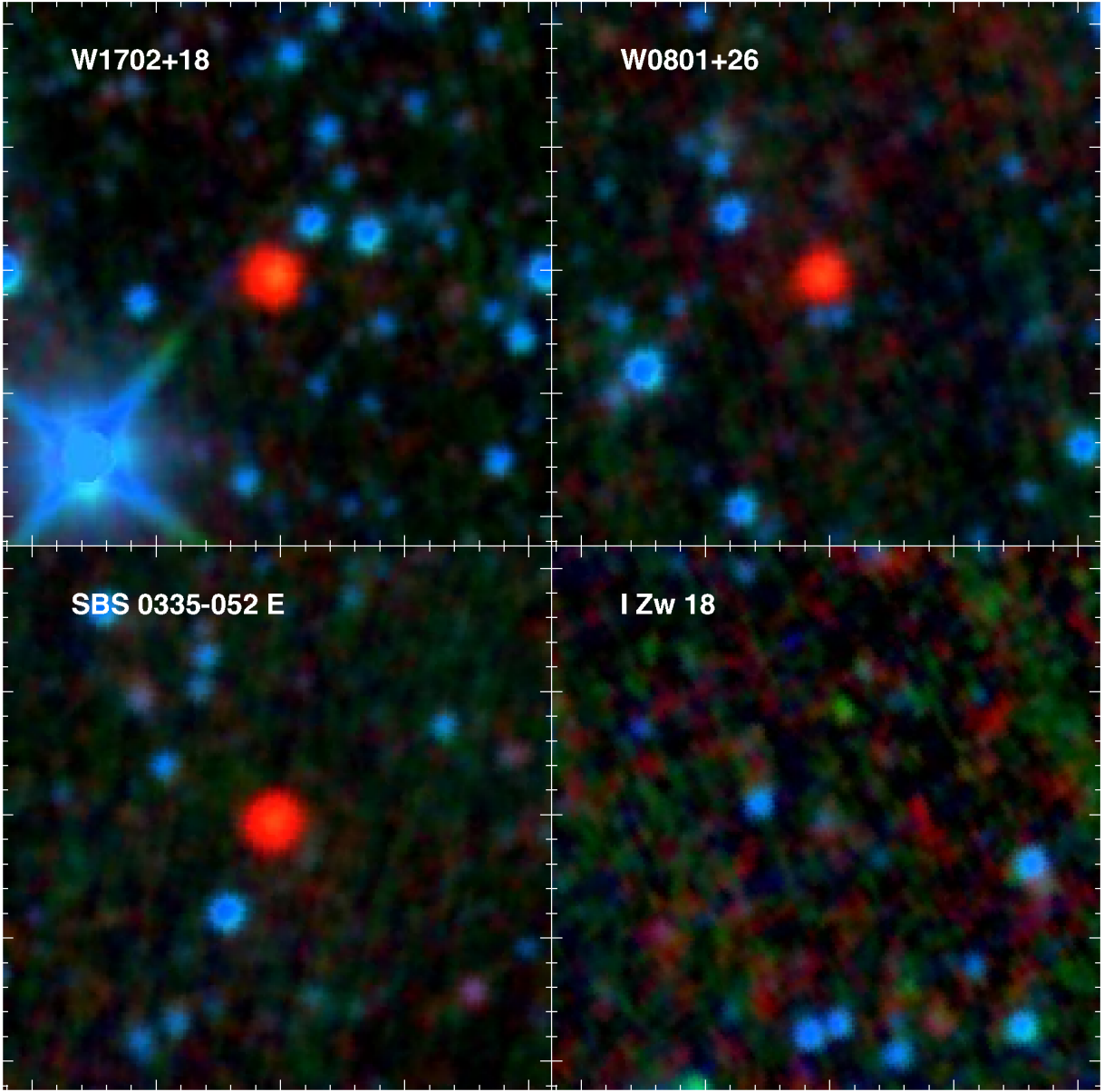}
\caption{{\it WISE} color images (blue = $W1$, green =$W2$, red = $W3$) for the
two newly discovered low metallicity BCDs as well as the prototypical
BCDs \sbs\ and I~Zw~18.  Images are 5\arcmin\ on a side, with
North up and East to the left.  ``Active'' BCDs such as \sbs\ and
the two new BCDs have unique, extremely red {\it WISE} colors, making them readily
identifiable from the {\it WISE} all-sky survey.}
\label{fig:images}
\end{figure}

\begin{figure}
\plotone{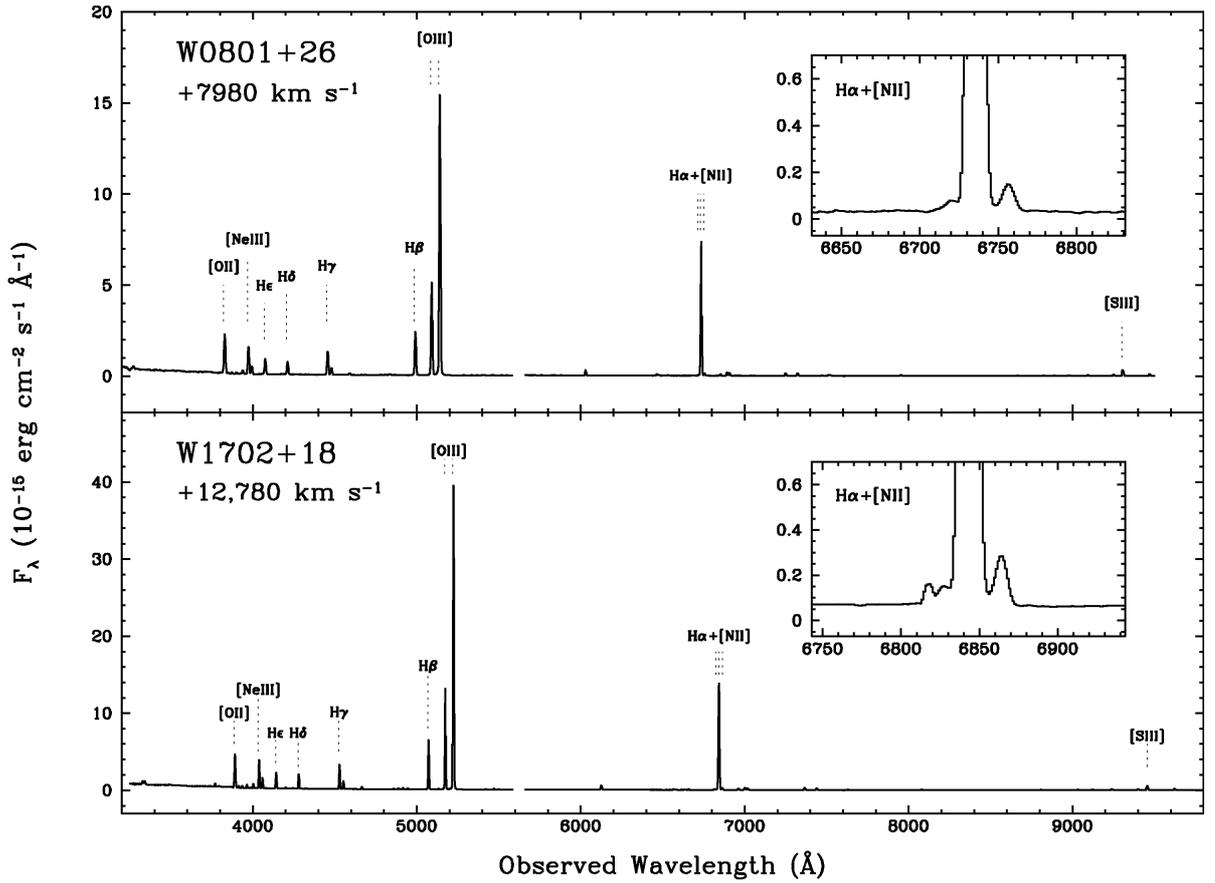}
\caption{Spectra of \wzero\ and \wone, obtained with the LRIS
instrument on Keck~I.  Both spectra show strong, narrow, high
equivalent width emission lines from hydrogen and oxygen, as well
as weak [\ion{N}{2}] flanking the H$\alpha$ emission line (insets).
The line ratios imply that the emission is powered by star formation,
not AGN activity, from very low metallicity systems.}
\label{fig:spectra}
\end{figure}

\begin{figure}
\plotone{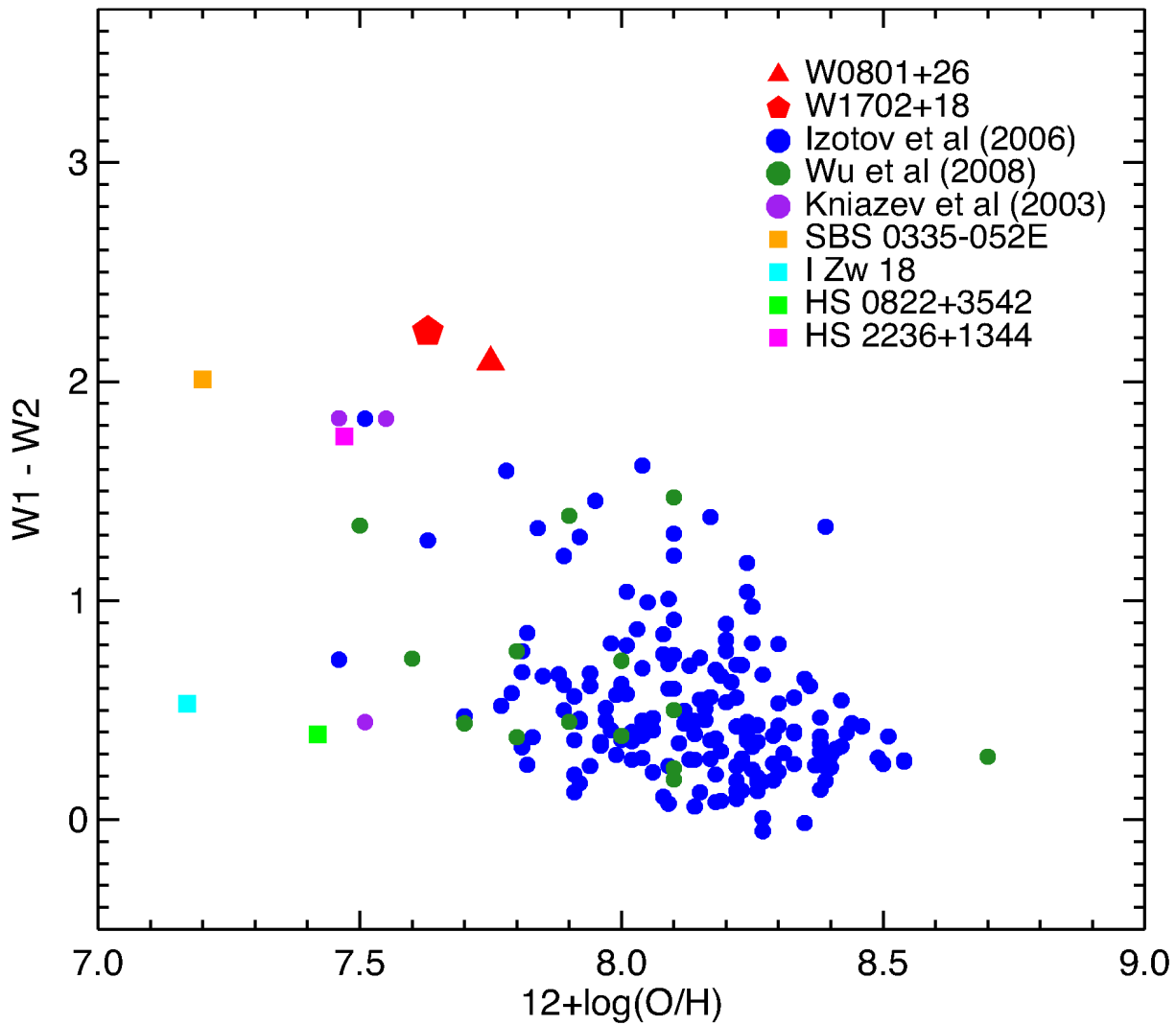}
\caption{Mid-infrared color (Vega) vs.  metallicity, $12+ \log{\rm (O/H)}$,
for several samples of BCDs, as indicated.  
The plotted sources are required to have SNR $>$ 5 in $W1$ and $W2$.  
For comparison, we also plot \sbs, I~Zw~18, HS~0822+3542 and HS 2236+1344,  four
of the lowest metallicity galaxies known to date.  \wzero\ and
\wone\ are significantly redder in $W1-W2$ than the majority of
known BCDs.}
\label{fig:metallicity_color}
\end{figure}

\eject

\clearpage

\begin{deluxetable}{lccccc}
\tablewidth{0pt}
\tablecaption{Properties of Low Metallicity BCDs.}
\tablehead{
\colhead{} &
\colhead{} &
\colhead{W0801+26} &  
\colhead{W1702+18} &
\colhead{SBS~0335-052E} &
\colhead{I~Zw~18}}
\startdata
R.A. (J2000) &&  08:01:03.929 &  17:02:33.534 &    03:37:44.032 &  09:34:02.088 \\
Dec. (J2000) && +26:40:53.91  & +18:03:06.44  & $-$05:02:40.26  & +55:14:26.23  \\
$z$          &&  0.0260 & 0.0425 &  0.0134 & 0.0025 \\
$M_B$ (mag)  &&  $-$16.8 & $-$18.3 & $-$17.1 & $-$15.4 \\
$B$ (mag)    && 18.18 & 18.39 & 16.48 & 14.46 \\
$W1$ (mag)   && 15.03 $\pm$ 0.04 & 14.26 $\pm$ 0.03 & 14.53 $\pm$ 0.03 & 15.39 $\pm$ 0.04 \\
$W2$ (mag)   && 12.94 $\pm$ 0.03 & 12.03 $\pm$ 0.02 & 12.52 $\pm$ 0.03 & 14.86 $\pm$ 0.07 \\
$W3$ (mag)   &&  8.29 $\pm$ 0.03 &  7.58 $\pm$ 0.02 &  7.66 $\pm$ 0.02 & 11.88 $\pm$ 0.14  \\
$W4$ (mag)   &&  5.38 $\pm$ 0.03 &  4.96 $\pm$ 0.03 &  5.04 $\pm$ 0.03 & 7.84 $\pm$ 0.14 \\
$12+\log{\rm (O/H)}$ &&  7.75$^{+0.04}_{-0.17}$  &  7.63 $\pm$ 0.06   & 7.11$-$7.3 & 7.17 \\
\enddata
\end{deluxetable}

\begin{deluxetable}{lcc}
\tablewidth{0pt}
\tablecaption{Spectroscopic Line Measurements.}
\tablehead{
\colhead{} &
\colhead{\wzero} &  
\colhead{\wone}}
\startdata
[\ion{O}{2}] 3727+3729	&  234.1 $\pm$ 2.6 & 334.2 $\pm$ 11.1  \\ 

H$\delta$ 4101		&  69.5 $\pm$ 1.1 &127.9 $\pm$ 2.0  \\

H$\gamma$ 4340		&  132.5 $\pm$ 4.1 &232.1 $\pm$ 13.7  \\

[\ion{O}{3}] 4363	&  41.5 $\pm$ 3.1 & 84.2 $\pm$ 8.7  \\

H$\beta$ 4861		&  247.3 $\pm$ 3.0 &438.5 $\pm$ 5.7  \\

[\ion{O}{3}] 4959	&  527.8 $\pm$ 11.3 & 921.8 $\pm$ 31.6 \\

[\ion{O}{3}] 5007	& 1596.4 $\pm$ 18.9 & 2679.2 $\pm$ 62.3  \\

H$\alpha$ 6563 	& 657.4 $\pm$ 9.0 &   1227.2 $\pm$ 27.5  \\

[NII] 6584			& 10.0   $\pm$ 0.8 &     21.1  $\pm$ 0.6 \\
                        &           & \\
$-$EW(H$\beta$)           & 370.2 $\pm$ 6.1 &     324.4 $\pm$ 6.4 \\
\enddata
\tablecomments{The top rows present the line fluxes in units of 10$^{-16}$ erg cm$^{-2}$
s$^{-1}$.  The final row gives the H$\beta$ equivalent width (EW) in units of \AA.}
\end{deluxetable}


\begin{thebibliography}{37}
\expandafter\ifx\csname natexlab\endcsname\relax\def\natexlab#1{#1}\fi

\bibitem[{{Allende Prieto} {et~al.}(2001){Allende Prieto}, {Lambert}, \&
  {Asplund}}]{AP01}
{Allende Prieto}, C., {Lambert}, D.~L., \& {Asplund}, M. 2001, \apjl, 556, L63

\bibitem[{{Baldwin} {et~al.}(1981){Baldwin}, {Phillips}, \&
  {Terlevich}}]{baldwin81}
{Baldwin}, J.~A., {Phillips}, M.~M., \& {Terlevich}, R. 1981, \pasp, 93, 5

\bibitem[{{Burgasser} {et~al.}(2011)}]{burgasser2011}
{Burgasser}, A. {et~al.} 2011, \apj, in press

\bibitem[{{Calzetti} {et~al.}(2007)}]{calzetti07}{Calzetti}, D. {et~al.} 2007, \apj, 666, 870

\bibitem[Cardelli et al.(1989)]{CCM89} Cardelli, J.~A., Clayton, G.~C., \& Mathis, J.~S.\ 1989, \apj, 345, 245 

\bibitem[{{Dale} {et~al.}(2001{\natexlab{a}}){Dale}, {Helou}, {Neugebauer},
  {Soifer}, {Frayer}, \& {Condon}}]{dale01}
{Dale}, D.~A., {Helou}, G., {Neugebauer}, G., {Soifer}, B.~T., {Frayer}, D.~T.,
  \& {Condon}, J.~J. 2001{\natexlab{a}}, \aj, 122, 1736

\bibitem[{{Dale} {et~al.}(2001{\natexlab{b}})}]{dale2001}
{Dale}, D.~A. {et~al.} 2001{\natexlab{b}}, \aj, 122, 1736

\bibitem[{{Davidson} \& {Kinman}(1985)}]{david85}
{Davidson}, K. \& {Kinman}, T.~D. 1985, \apjs, 58, 321

\bibitem[{{Dufour} \& {Hester}(1990)}]{dufour90}
{Dufour}, R.~J. \& {Hester}, J.~J. 1990, \apj, 350, 149

\bibitem[{{Engelbracht} {et~al.}(2008){Engelbracht}, {Rieke}, {Gordon},
  {Smith}, {Werner}, {Moustakas}, {Willmer}, \& {Vanzi}}]{Engelbracht08}
{Engelbracht}, C.~W., {Rieke}, G.~H., {Gordon}, K.~D., {Smith}, J., {Werner},
  M.~W., {Moustakas}, J., {Willmer}, C.~N.~A., \& {Vanzi}, L. 2008, \apj, 678,
  804

\bibitem[{{Hirashita} \& {Hunt}(2004)}]{hirash04}
{Hirashita}, H. \& {Hunt}, L.~K. 2004, \aap, 421, 555

\bibitem[{{Houck} {et~al.}(2004)}]{houck04}
{Houck}, J.~R. {et~al.} 2004, ApJS, 154, 211

\bibitem[{{Izotov} {et~al.}(1990){Izotov}, {Guseva}, {Lipovetskii}, {Kniazev},
  \& {Stepanian}}]{izotov90}
{Izotov}, I.~I., {Guseva}, N.~G., {Lipovetskii}, V.~A., {Kniazev}, A.~I., \&
  {Stepanian}, J.~A. 1990, in Astrophysics and Space Science Library, Vol. 162,
  Physical Processes in Fragmentation and Star Formation, ed.
  {R.~Capuzzo-Dolcetta, C.~Chiosi, \& A.~di Fazio}, 235--239

\bibitem[{{Izotov} {et~al.}(1997){Izotov}, {Lipovetsky}, {Chaffee}, {Foltz},
  {Guseva}, \& {Kniazev}}]{izotov97}
{Izotov}, Y.~I., {Lipovetsky}, V.~A., {Chaffee}, F.~H., {Foltz}, C.~B.,
  {Guseva}, N.~G., \& {Kniazev}, A.~Y. 1997, \apj, 476, 698

\bibitem[{{Izotov} {et~al.}(2006){Izotov}, {Stasi{\'n}ska}, {Meynet}, {Guseva},
  \& {Thuan}}]{izotov06}
{Izotov}, Y.~I., {Stasi{\'n}ska}, G., {Meynet}, G., {Guseva}, N.~G., \&
  {Thuan}, T.~X. 2006, \aap, 448, 955

\bibitem[{{Izotov} \& {Thuan}(2007)}]{izotov2007}
{Izotov}, Y.~I. \& {Thuan}, T.~X. 2007, \apj, 665, 1115

\bibitem[{{Kewley} {et~al.}(2001){Kewley}, {Dopita}, {Sutherland}, {Heisler},
  \& {Trevena}}]{kewly01}
{Kewley}, L.~J., {Dopita}, M.~A., {Sutherland}, R.~S., {Heisler}, C.~A., \&
  {Trevena}, J. 2001, \apj, 556, 121

\bibitem[{{Kirkpatrick} {et~al.}(2011)}]{kirkpatrick2011}
{Kirkpatrick}, J.D. {et~al.} 2011, \apj, submitted

\bibitem[{{Kniazev} {et~al.}(2003){Kniazev}, {Grebel}, {Hao}, {Strauss},
  {Brinkmann}, \& {Fukugita}}]{knia03}
{Kniazev}, A.~Y., {Grebel}, E.~K., {Hao}, L., {Strauss}, M.~A., {Brinkmann},
  J., \& {Fukugita}, M. 2003, \apjl, 593, L73

\bibitem[{{Kunth}(1999)}]{kunth99}
{Kunth}, D. 1999, \apss, 265, 489

\bibitem[{{Kunth} {et~al.}(1994){Kunth}, {Lequeux}, {Sargent}, \&
  {Viallefond}}]{kunth94}
{Kunth}, D., {Lequeux}, J., {Sargent}, W.~L.~W., \& {Viallefond}, F. 1994,
  \aap, 282, 709

\bibitem[{{Kunth} \& {{\"O}stlin}(2000)}]{kunth2000}
{Kunth}, D. \& {{\"O}stlin}, G. 2000, \aapr, 10, 1

\bibitem[{{Kunth} \& {Sargent}(1986)}]{kunth86}
{Kunth}, D. \& {Sargent}, W.~L.~W. 1986, \apj, 300, 496

\bibitem[{{Lauqu{\'e}}(1973)}]{lauqu73}
{Lauqu{\'e}}, R. 1973, \aap, 23, 253

\bibitem[{{Mainzer} {et~al.}(2011)}]{mainzer2011}
{Mainzer}, A. {et~al.} 2011, \apj, 726, 30

\bibitem[{{Massey} \& {Gronwall}(1990)}]{massey90}
{Massey}, P. \& {Gronwall}, C. 1990, \apj, 358, 344

\bibitem[Meier 
\& Terlevich(1981)]{meier81} Meier, D.~L., \& Terlevich, R.\ 1981, \apjl, 246, L109 

\bibitem[{{Melnick} {et~al.}(1992){Melnick}, {Heydari-Malayeri}, \&
  {Leisy}}]{mel92}
{Melnick}, J., {Heydari-Malayeri}, M., \& {Leisy}, P. 1992, \aap, 253, 16

\bibitem[{{Oke} {et~al.}(1995)}]{oke95}
{Oke}, J.~B. {et~al.} 1995, \pasp, 107, 375

\bibitem[{{Schaerer} \& {Vacca}(1998)}]{schaerer98}
{Schaerer}, D. \& {Vacca}, W.~D. 1998, \apj, 497, 618

\bibitem[{{Schlegel} {et~al.}(1998){Schlegel}, {Finkbeiner}, \&
  {Davis}}]{schlegel98}
{Schlegel}, D.~J., {Finkbeiner}, D.~P., \& {Davis}, M. 1998, \apj, 500, 525

\bibitem[Smith 
\& Hancock(2009)]{smith09} Smith, B.~J., \& Hancock, M.\ 2009, \aj, 138, 130 

\bibitem[{{Thuan} {et~al.}(1997){Thuan}, {Izotov}, \& {Lipovetsky}}]{thuan97}
{Thuan}, T.~X., {Izotov}, Y.~I., \& {Lipovetsky}, V.~A. 1997, \apj, 477, 661

\bibitem[{{Thuan} \& {Martin}(1981)}]{thuan81}
{Thuan}, T.~X. \& {Martin}, G.~E. 1981, \apj, 247, 823

\bibitem[{{Thuan} {et~al.}(1999){Thuan}, {Sauvage}, \& {Madden}}]{thuan99}
{Thuan}, T.~X., {Sauvage}, M., \& {Madden}, S. 1999, \apj, 516, 783

\bibitem[{{Wright} {et~al.}(2010)}]{wright10}
{Wright}, E.~L. {et~al.} 2010, \aj, 140, 1868

\bibitem[{{Wu} {et~al.}(2007){Wu}, {Charmandaris}, {Hunt}, {Bernard-Salas},
  {Brandl}, {Marshall}, {Lebouteiller}, {Hao}, \& {Houck}}]{Wu07}
{Wu}, Y., {Charmandaris}, V., {Hunt}, L.~K., {Bernard-Salas}, J., {Brandl},
  B.~R., {Marshall}, J.~A., {Lebouteiller}, V., {Hao}, L., \& {Houck}, J.~R.
  2007, \apj, 662, 952

\bibitem[{{Wu} {et~al.}(2008)}]{wu08}
{Wu}, Y. {et~al.} 2008, \apj, 676, 970

\bibitem[{{Zwicky}(1966)}]{zwicky66}
{Zwicky}, F. 1966, \apj, 143, 192

\end{thebibliography}
\end{document}